\documentclass[
    ,final            
  ]
  {aipproc}

\layoutstyle{6x9}

\begin{document}

\title{Recent Physics Results with the COMPASS Experiment}

\author{Stephan Paul\\{\small for the COMPASS Collaboration}}
{ address={Technische Universit\"{a}t M\"{u}nchen, Physik Department
E18\newline James-Franck Strasse, D-85478 Garching} }


\begin{abstract}
The COMPASS experiment has obtained first physics results in the
field of polarized distribution functions for quarks and gluons
using muon scattering off polarized deuterons. The analysis using
open charm production and pairs of high $p_T$ hadrons is presented.
We also have used a transversely polarized target to address
transverse information for quarks inside the nucleon. In addition, a
pilot run with incoming pions taken late 2004 will give first
information on the pion polarizabilities and hadron resonances. The
physics prospects from this run as well as from future data taking
in this field are also outlined.

\end{abstract}
\maketitle


\section{Introduction}

The structure of hadrons can be investigated at different length
scales revealing different properties and descriptions of hadrons.
At small distance scales (large Q$^{2}$), deep inelastic scattering
\ reveals the nucleon structure in terms of current quarks and
gluons, using structure functions. The hot topic are spin degrees of
freedom, the understanding of the nucleons spin in the interplay of
its constituents. At medium distance scales the tool of spectroscopy
is useful, a domain still dominated by the language of constituent
quarks. The search for exotic quantum numbers and glueballs\ at
masses between 1.5 and 2.5 GeV/c$^{2}$ should reveal more insights
to the phenomenon of confinement. Very large distance scales probe
quasi static hadron properties like polarizabilities of mesons.
Effective field theories derived from QCD are expected to give a
consistent description of the low energy quantities. The COMPASS
experiment at CERN offers the unique opportunity of studying the
three fields of hadron physics within one experiment using flexible
beams and spectrometers setups. As most of the beam time has so far
been devoted to experiments using muons (deep inelastic scattering -
DIS and hard interactions) most of this report will be devoted to
first results on this topic. At the end of last year a first pilot
run with pions has been performed aiming at the measurement of
polarizabilities and diffractive meson production. A first glimpse
to this program will end this report.

\section{The COMPASS Experiment}
\begin{figure}
  \includegraphics[width=3.55in]{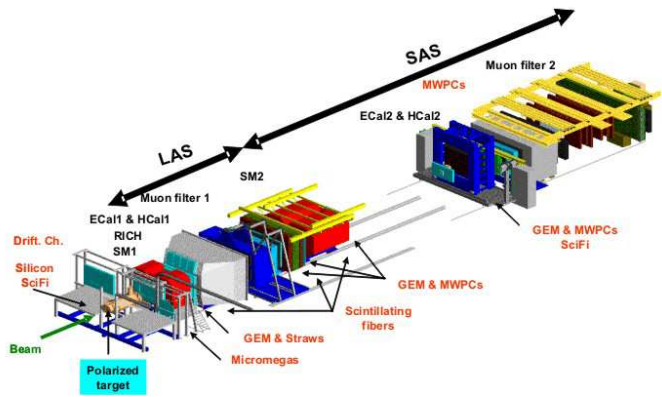}\\
  \caption{Layout of the COMPASS experiment at the CERN SPS}\label{COMPASS}
\end{figure}

The COMPASS experiment at the CERN SPS is a modern 2-stage magnetic
spectrometer with a flexible setup to allow for a variety of physics
programs to be performed with different beams (fig. \ref{COMPASS}).
Common to all measurements is the requirement for high beam
intensity and interaction rates with the needs of a high readout
speed. The spectrometer is equipped with tracking systems based on
silicon detectors for high precision tracking in the target region,
micromega and GEM detectors for small area and wire/drift chambers
as well as straw tubes for large area tracking. Particle
identification is performed in a large acceptance ring imaging
cherenkov detector downstream of the first spectrometer magnet and
in hadronic (HCAL) and electromagnetic (ECAL) calorimeters in both
parts of the spectrometer. The calorimeters are also used for the
energy reconstruction of neutral particles. Both parts of the
spectrometer are also equipped with muon detectors.
\par
For measurements with a polarized nucleons a polarized target is
installed about 3m upstream of the first magnet. It consists of two
independent target cells filled with $^6LiD$ crystals mounted inside
a very homogeneous solenoidal field of 2T. Using the technique of
dynamic nuclear polarization 50\% of the nucleons could be polarized
with a  degree polarization of more than 50\%. The two target cells
are thereby operated with opposite polarization to reduce systematic
effects on asymmetry measurements owing to time dependent
performance of the apparatus. The study of exclusive reactions using
hadron beams requires a short solid target inside a veto-detector
surrounded by silicon telescopes. The veto-detector is sensitive to
charge and neutral particles with small openings in forward and
backward direction. \linebreak

The trigger system is mostly based on scintillator hodoscopes
mounted in front of the target and downstream of the second magnet.
Using a muon beam deep inelastic scattering (DIS) at large $Q^2$ is
selected as well as low $Q^2$ events (quasi-real photons) with
accompanying hadrons observed in HCAL. For diffractive and Primakoff
scattering with hadron beams the target is surrounded by a veto box
and small multiplicity events are selected by means of a
scintillator and within an online event filter operated inside the
event building computers.
\section{Physics results with muon
beams}

\subsection{Spin-structure functions}

\bigskip The low energy description of the nucleons spin is based on the SU(6)
wave functions for non-interacting constituent quarks of spin
$\frac{1}{2}$ where two quarks are spin-aligned and one quark
spin-antialligned making a total spin of $\frac{1}{2}$ for the
nucleon. In turn, at small distance scales the nucleon is described
in terms of QCD where quarks interact among each other via gluons.
The resulting sharing of the nucleons momentum is described via
structure functions (number densities) which are a function of the
Bjorken variable x denoting the momentum fraction of a constituent
and the energy scale Q$^{2}$ (which sometimes is identified with a
spatial resolution scale). Quarks and gluons are not independent
partners and the repartitioning of kinematic variables when changing
the energy scale is described by the DGLAP equations. Measurements
of the quark helicity within a polarized nucleon has been performed
by a number of experiments using DIS on polarized nucleons at
various $Q^{2}$. The pictorial interpretation of the results are not
straightforward but its safe to say that quarks only carry part of
the nucleons spin. Owing to the axial anomaly (coupling of quarks
and gluons) the singlet axial charge $a_0$ (which enters the cross
section asymmetries) combines spin contributions of quarks
$\Delta\Sigma$ and gluons $\Delta G$

\begin{equation}
a_{0}=\Delta\Sigma-\frac{3\alpha_{a}\left(  Q^{2}\right)  }{2\pi
}\Delta G
\label{axial_singlet_charge}
\end{equation}
\begin{equation}
{\textnormal with} ~\Delta\Sigma=\Delta u+\Delta d+\Delta
s=2\left\langle S_{z} ^{quark}\right\rangle \label{Delta_sigma}
\end{equation}

Using a QCD analysis of the data we can extract values for $\Delta
\Sigma\sim 0.2$. The accuracy is limited by the data quality at very
low values of x, as $\Delta\Sigma$ is obtained from the integral of
the spin-dependent quark structure function $g_{1}(x)$. The same
analysis also allows to extract an estimate of the gluon spin
structure-function $\Delta G(x)$. As experiments probe $a_{0}$ at
different values of $Q^{2}$ we can use the evolution equations to
obtain an estimate of the gluon contribution of the spin. However,
the $Q^{2}$ range is small and thus uncertainties on $\Delta G(x)$
are large.

\subsubsection{Measurement of $A_{1}$ on the deuteron}

\begin{center}
\begin{figure}
\includegraphics[width=2.1811in ] {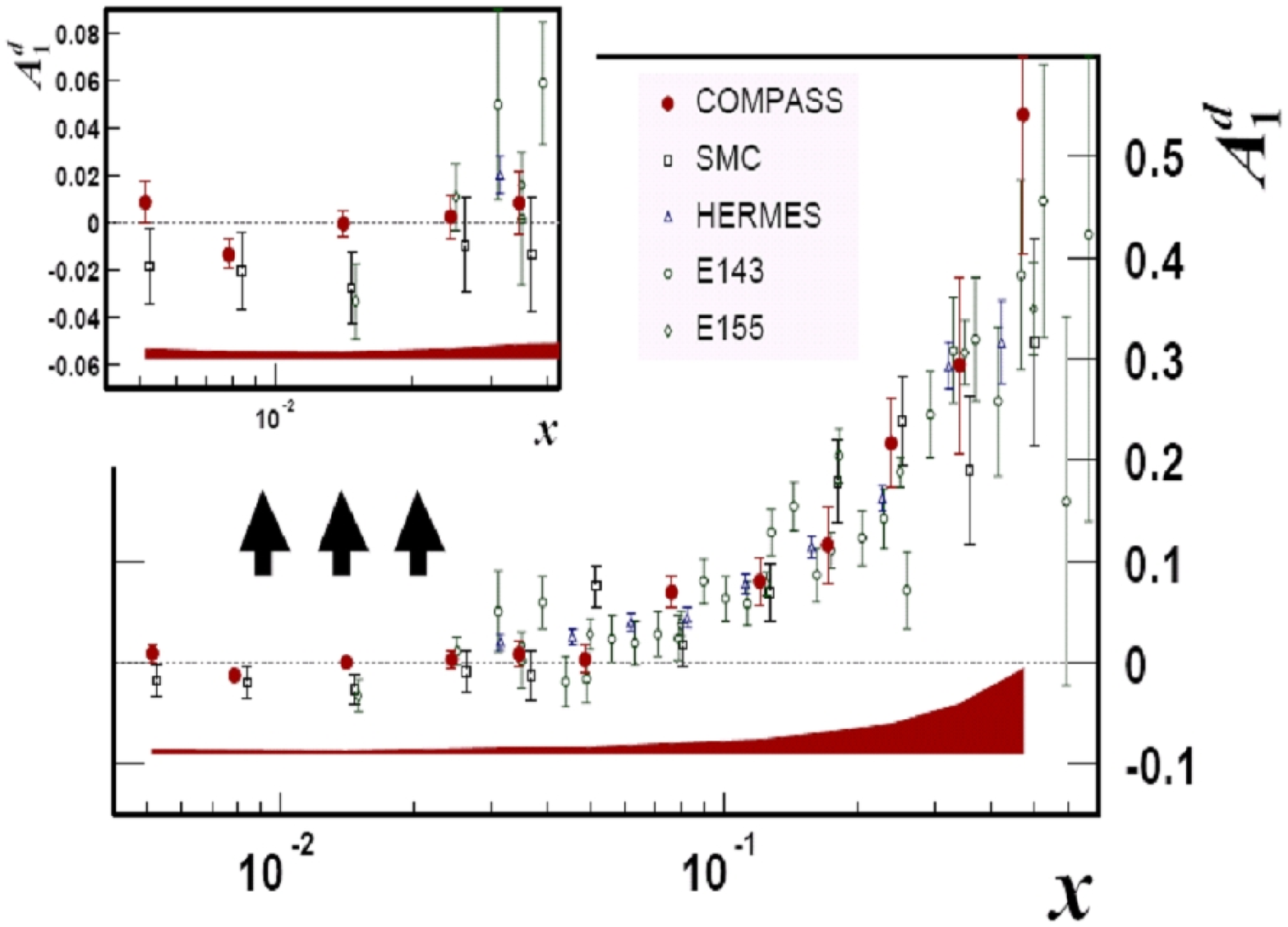}
\includegraphics[width=2.1603in ] {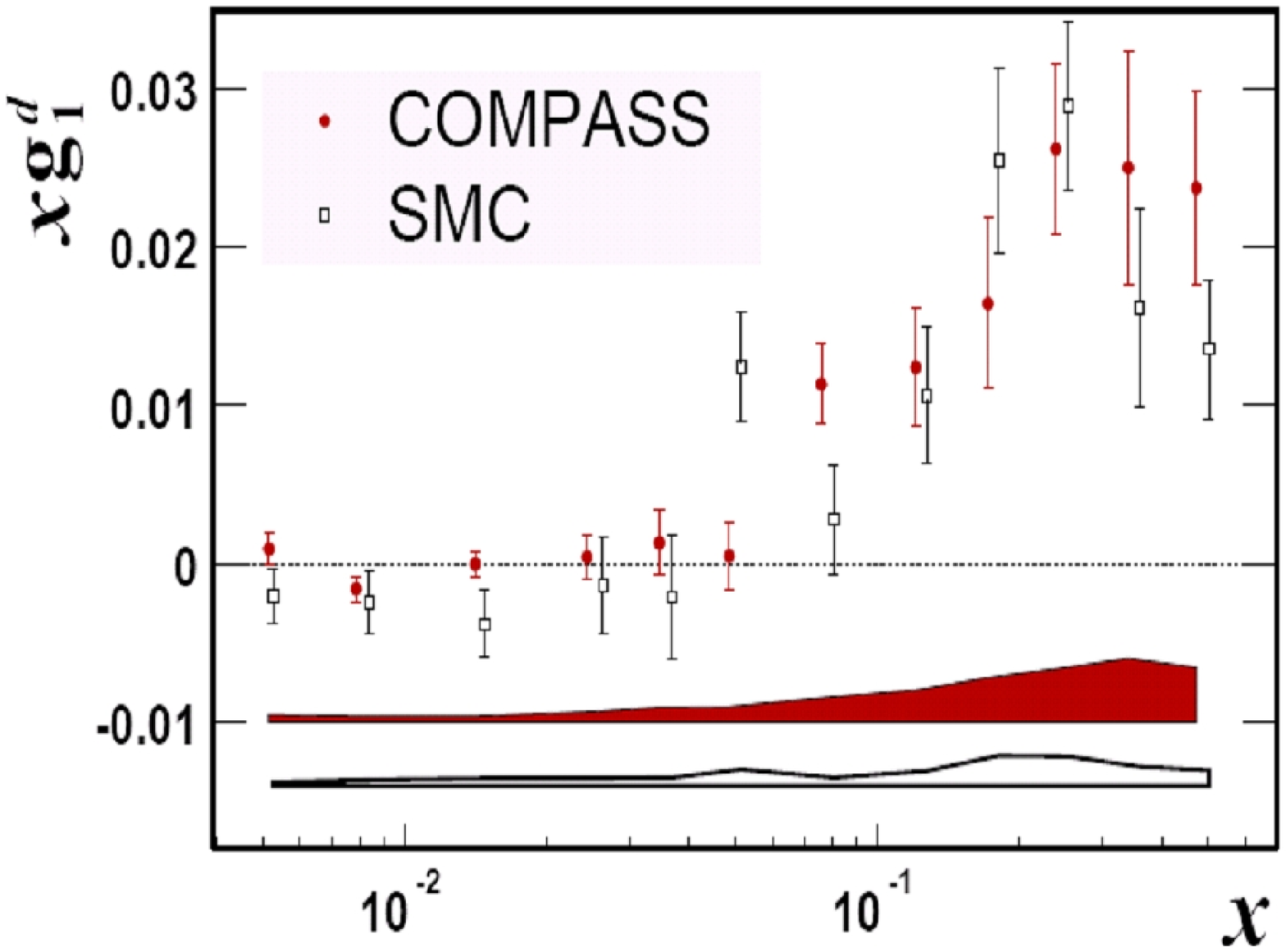} \caption{Left: Measured asymmetry
using DIS events on the deuteron. The data are compared to results
from previous experiments. The area shaded in dark red denotes the
systematic uncertainties. The inlet depicts the results for small
values of x, where COMPASS has the largest impact. Right: Results
for $x\cdot g_{1}(x)$ extracted from the measured asymmetry. Data
are taken at large $Q^{2}$ and compared to results from the SMC
experiment. Bands for systematic uncertainties are given for both
data sets (COMPASS: dark red)} \label{A1_plot}
\end{figure}
\end{center}

The high rate capability of COMPASS allows a precision study of the
quark asymmetry $A_{1}^{d}$ on the deuteron \cite{COMPASS_A1}. Using
data taken in '02 and '03 we have collected about $3\cdot10^{7}$
DIS-events. The measured asymmetry $A_{1}^{d}$ is obtained combining
data from periods with opposite spin polarizations within the two
target cells (see fig. \ref{A1_plot} left). Thus, acceptance effects
cancel out. We can translate the asymmetries into the polarized
quark distribution functions normalizing the measured asymmetry to
the unpolarized cross section determined by the structure function
$F_2(x)/2x$ and the ratio of longitudinal to transverse cross
section R.
\begin{equation}
g_{1}^{d}=\frac{F_{2}^d}{2x(1+R)}A_{1}^{d}
\end{equation}

The results for $x\cdot g_{1}(x)$ extracted from the measured
asymmetry are shown in the right of fig. \ref{A1_plot}. The data are
taken at large $Q^{2}$ and are compared to results from the SMC
experiment. Systematic uncertainties are marked by bands and are
given for both data sets (COMPASS: dark red). The high precision of
the new COMPASS data at low x help to better define the integral
over $g_{1}(x)$ and thus allow to reduce the uncertainties on
$\Delta\Sigma$ by a factor 2.
\begin{equation}
\Delta\Sigma=\ 0.237_{-0.029}^{+0.024}
\end{equation}

\subsubsection{The Gluon Spin Contribution}

From the results obtained for the spin contribution of the quarks we
conclude that the picture must be far more complicated than assumed
at first. We can thus write the nucleons spin formally as a sum of
four contribution.
\begin{equation}
\frac{1}{2}=S_{z}=\frac{1}{2}\Delta\Sigma+\Delta
G+L_{q}+L_{G}\label{spin_sum}
\end{equation}
However, the interpretation and distinction between the different
terms depends on the reference system and thus is not unique. While
contributions to angular momenta are currently only accessible in
principle\footnote{In the recent years a new mathematical
formulation of angular momentum inside the nucleons in terms of so
called generalized parton distributions has been found. However,
this requires a large set of exclusive measurements over a wide
kinematic range for which dedicated experiments and machines will
have to be constructed. It turns out that also the measurement of
transversity inside the nucleon is connected to angular momentum via
the determination of the Sivers function. However, this will be very
model dependent.} the COMPASS experiment aims at determining the
role of the gluon spin to this sum. This requires scattering
processes which are a) sensitive to gluons and b) depict a sizeable
analyzing power. In COMPASS the photon-gluon fusion process (PGF) is
used, which can be tagged in two different ways.

\begin{enumerate}
\item The production of open charm is a very clean sign for the underlying
hard QCD process as charm production in either fragmentation
processes or via intrinsic charm is negligible. The energy scale of
the process is determined by the minimal value
$\sqrt{s}\geq2m_{c}\simeq3GeV/c^{2}$, thus we can use quasi real
photons ($Q^{2}\simeq0$).
\item A second tag is the pair production of light hadrons at large transverse
momentum ($p_{T}\geq1GeV/c$) which also determines the scale of the
underlaying process.
\end{enumerate}

\noindent
We will discuss the results for both processes studied.

\subsubsection{$\Delta G$ via production of open charm}

\begin{center}
\begin{figure}[ptb]
\includegraphics[width=2.2451in ] {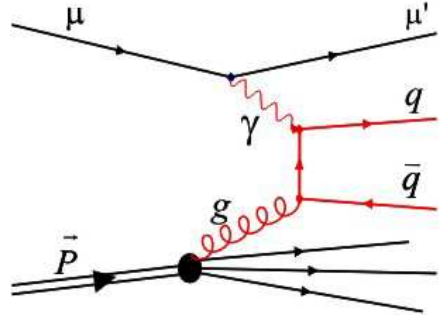} \caption{The $\gamma
g$-fusion process. It can be tagged using open charm or high
$p_{T}$-hadron pairs in the final state.} \label{photon_gluon}
\end{figure}
\end{center}

\begin{figure}[ptb]
\includegraphics[
width=2.4293in ] {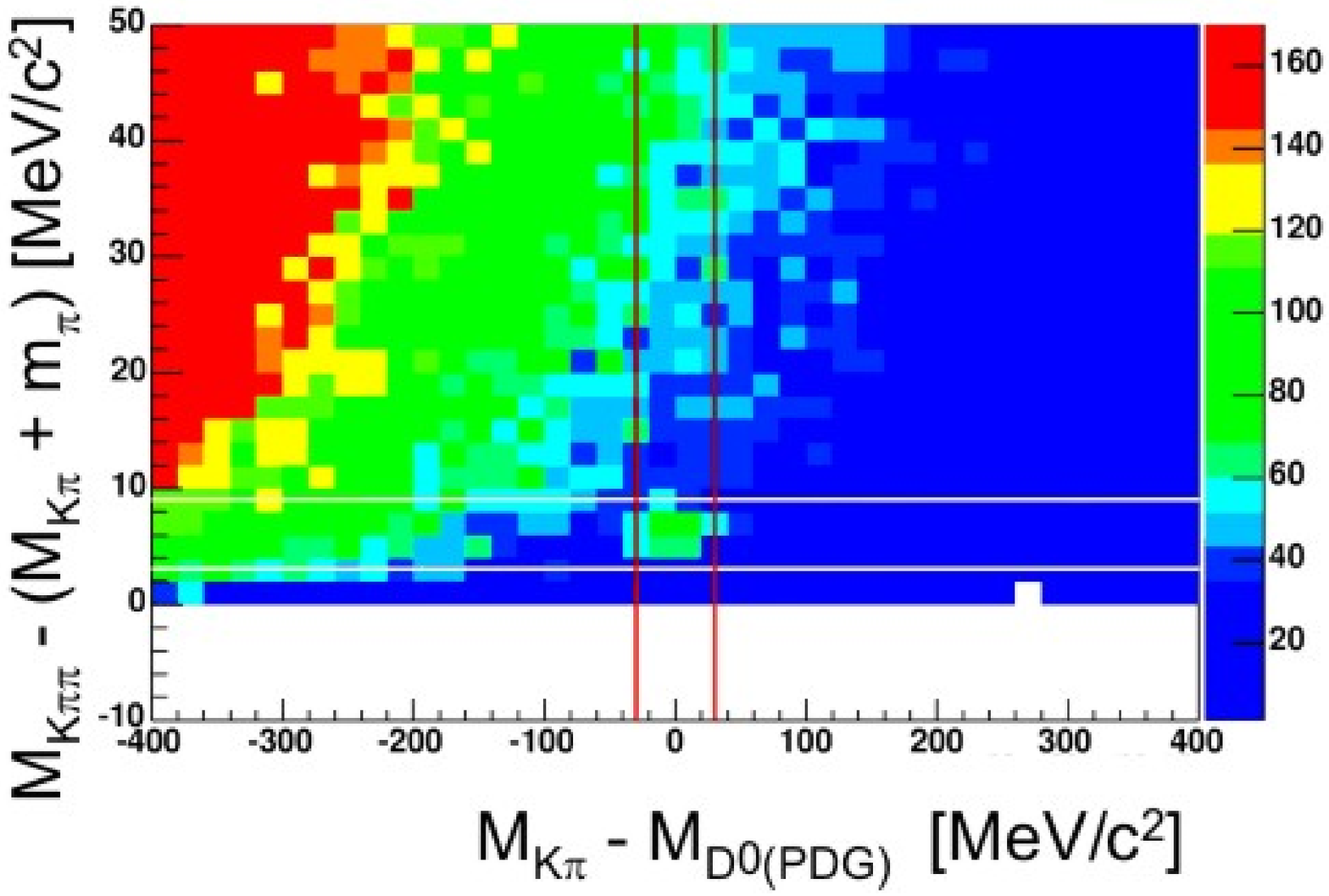} \hspace{.5cm}
\includegraphics[
width=2.5108in ] {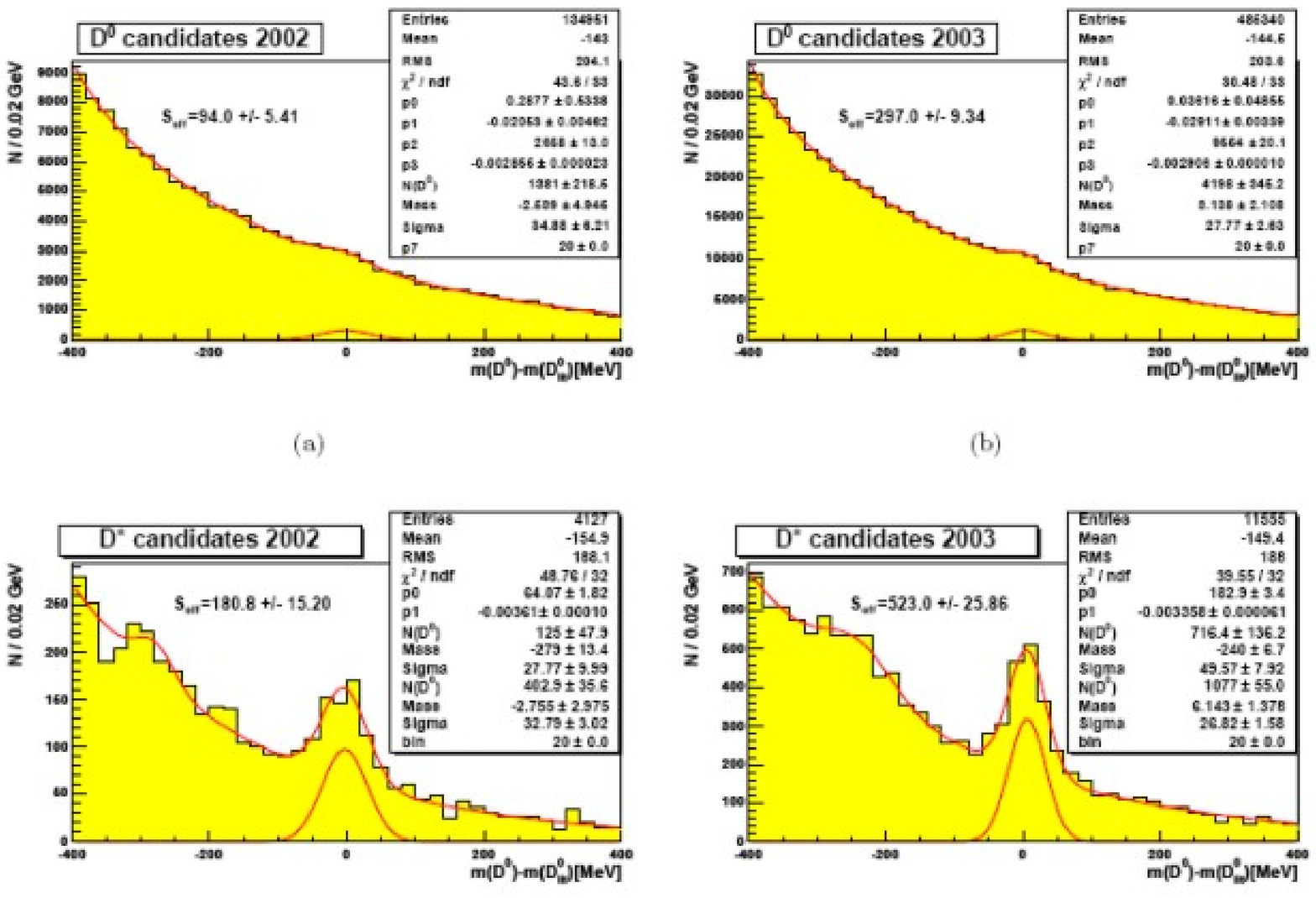} \caption{Left:
~$M(K\pi\pi)-M(K\pi )$ versus the $M(K\pi )-M(D^{0})$ mass
difference. Right: ($K\pi$) invariant mass plots showing
$D^{0}$-signals for two different selection schemes and the two data
periods (left: 2002, right: 2003). Upper plots:$(K\pi )$ with hard
K-identification. Lower plots:$(K\pi )$ combinations tagged by slow
pions consistent with $D^{\ast}-D^{0}$ mass difference.}
\label{D_mass_plots}
\end{figure}

The PGF-process is depicted in fig. \ref{photon_gluon}. The yield of
charmed hadrons is about 1\%, the analyzing power 10\%. Using data
obtained in 2002 and 2003 we have extracted a sample of D-mesons.
This is done in two ways. $D^{0}$ candidates are selected via there
decay $D^{0}\rightarrow K\pi$ where the kaons are identified within
the RICH. A very clean signal is obtained using reconstructed
$D^{\ast}$-decays ($D^{\ast }\rightarrow D^{0}\pi_{slow}$). These
events are shown on the left of fig. \ref{D_mass_plots}. The full
$D^{0}$ sample with and without $D^{\ast}$-tagging is summarized on
the right of fig. \ref{D_mass_plots} From these data we can extract
$\Delta G$ using NLO calculations for the spin dependence of the
$\gamma g-$process.The result is $\frac{\Delta G}{G}=-1.08\pm0.73$
at a mean value of $\left\langle x_{G}\right\rangle =0.15$. We
expect to improve the statistical error by about a factor two using
2004 data.

\subsubsection{$\Delta G$ via production of high $p_{T}$ hadron pairs}

\begin{center}
\begin{figure}
\includegraphics[
width=2.7608in ] {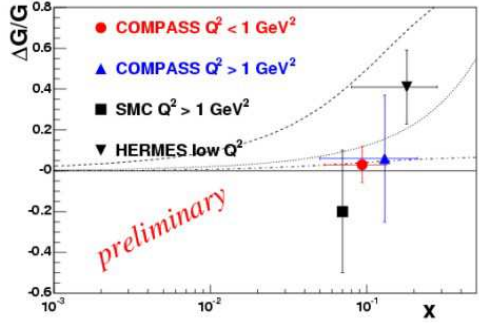} \caption{World data on direct
measurements of the gluon polarization $\Delta G/G$ using high $p_T$
hadron pairs \cite{COMPASS_pt}. The results from the COMPASS
analysis for both ranges of $Q^2$ are shown together with previous
analysis from SMC \cite{SMC_DeltaG} and HERMES \cite{HERMES_pt}.
Note that the HERMES result does not include a systematic
uncertainty linked to the background subtraction from other
processes but PGF. Experimental curves taken from
\cite{DeltaG_theory}} \label{delta_g_compass}
\end{figure}
\end{center}

The channel of high $p_{T}$ hadron pairs shares the virtue of high
statistics with the drawback of systematic uncertainties on the role
of background processes. The selection requires individual
$p_{T}\geq 0.7$ combined with
$p_{1T}^{2}+p_{2T}^{2}\geq(2.5(GeV/c)^{2}$ as well as
 $m(h_{1},h_{2})>1.5GeV/c^{2}$.
The data set is divided into $Q^2> 1(GeV/c)^{2}$ and $Q^2\leq
1(GeV/c)^{2}$, which in our data set corresponds to quasi real
photon production. For the high $Q^2$ sample the background
processes are ordinary DIS events with fragmentation or
gluon-Bremsstrahlung. However, as the quark asymmetries $A_1^d$ are
negligible at small x, these two processes do not dilute the
asymmetry from the underlaying PGF process for the analyzing power
and the partial cross section are taken from theory (and MC). The
result using data from 2002 and 2003 is:
\begin{equation}
\Delta G/G = 0.06 \pm 0.31 (stat.) \pm 0.06 (syst.) ~~for~Q^2> 1
(GeV/c)^{2}; <x_B>~1.15
\end{equation}
For quasi-real photons the main background processes in the
asymmetry are contributions from resolved photons (fluctuations of
the photon into a hadron). As the polarized structure function of
the photon is unknown, these processes contribute largely to the
systematic errors as the contribution is model dependent and maximal
and minimal scenarios are used to obtain a handle on their
contribution. It should be noted that all corrections are only
calculated in LO. From the measured asymmetry $A_{LL}/D=0.002
\pm0.019 (stat)$ we derive a gluon polarization of
\begin{equation}
\Delta G/G = 0.024 \pm 0.089 (stat) \pm 0.057 (syst) ~~for~Q^2\leq 1
(GeV/c)^{2}; <x_B>~0.95
\end{equation}
Fig. \ref{delta_g_compass} summarizes the COMPASS results for
$\Delta G/G$ together with previous measurements from SMC and
HERMES. It should be noted that previous determinations have not
used the same method of subtracting background which is due to other
processes. The small value of the gluon polarization at $x\sim 0.1$
can be compared to various models for the functional dependence
$\Delta G(x)/G(x)$ assuming different values for the integral value.
It seems that large values of $\Delta G/G$ are very unlikely.
\subsection{Transverse Spin Effects}
\begin{figure}[ht]
  \includegraphics[width=1.8in]{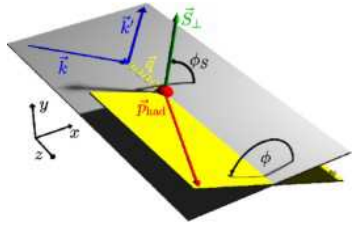} \hspace{.75cm}
  \includegraphics[width=2.7in]{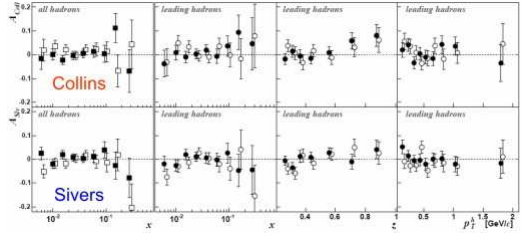}
  \caption{Left: Definition of angles in transversity measurements. Right: Azimuthal asymmetries obtained with a transversely polarized
target in COMPASS.
  Upper: Collins angles, lower: Sivers angle.}\label{transversity_figure}
\end{figure}

In addition to the spin distribution functions G(x) and g$_1$(x) for
longitudinally polarized nucleons the transversity function $\delta
q(x)$ describes the distribution of transversely polarized quarks
inside a transversely polarized nucleon. This function is more
difficult to measure and to interpret, though formally it has to be
considered on the same footing as the other structure functions. The
reason for the experimentally difficult access to this function is
that a physical measurement requires a suitable analyzing function.
One possibility is a spin dependent fragmentation function (Collins
function) in which a production asymmetry arises for hadron produced
with larger transverse momentum (eq. \ref{transversity_formula}).
This asymmetry has to be measured w.r.t. to the spin direction of
the nucleon and is called 'Collins function' \cite{Collins}. We
expect a modulation of the hadron intensity with
$\sin(\phi+\phi_s)$. The plane of reference for all azimuth angles
is the lepton scattering plane (see fig. \ref{transversity_figure}
for the definition of angles).
\begin{equation}
    A_T\propto \delta q(x)\cdot H_1(z,k_T)
    \label{transversity_formula}
\end{equation}
HERMES has measured the combination of the two functions for the
first time on polarized protons and obtained a non-zero result
pointing to both functions being non zero
\cite{HERMES_transversity}. The results for COMPASS measured on
transversely polarized deuterons \cite{COMPASS_transversity} are
depicted in fig. \ref{transversity_figure}.
\noindent A second effect leading to non-zero asymmetries in
reactions on transversely polarized targets is the Sivers effect
\cite{Sivers} which correlates transverse momentum of quarks inside
the nucleon with the transverse spin direction of the nucleon
$\sin(\phi-\phi_s)$. Sivers and transversity/Collins effects can
thus be disentangled by their different azimuthal dependence.

The COMPASS results show no sign of any asymmetry which might be
explained by a cancelation of the effect by neutron and proton
inside the deuteron.

\section{Diffractive Processes and Hadron Structure}
Using an incoming hadron beam COMPASS aims at studying hadron
polarizabilities using Primakoff reactions and light meson
spectroscopy in diffractive and central production. In either case
high rates and good acceptance should allow to improve over the
existing data sets by about a factor 10 in the statistical sample.

\subsection{A pilot run with Pions}
\begin{figure}[hb]
  \includegraphics[height=1.1in, width=1.5in]{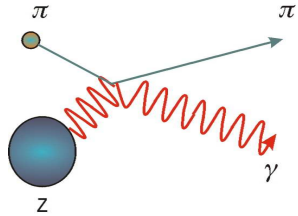}\\
  \caption{Sketch of the Primakoff reaction}\label{primakoff}
\end{figure}

In 2004 COMPASS has performed a short run with low intensity pions
($2\cdot10^6 \pi/spill$) impinging on a segmented lead (copper)
target of 50\% and 25\% radiation length (two modes of running). The
main purpose was as measurement of the pion polarizability using
scattering off virtual photons on a lead nucleus. The Compton
scattering process in inverse kinematics (fig. \ref{primakoff}) is
dominated by the Thompson cross section for point like particles and
contains two terms describing the electric and magnetic
polarizabilities of the hadron which can be separated by their
different angular dependence (see eq. \ref{Primakoff}, where the
cross section is given in the reference frame of the outgoing pion).
\begin{equation}\label{Primakoff}
    \frac{d\sigma_{\gamma\pi}(\theta,\omega)}{d\omega d\theta}=\frac{2\alpha^2}{m_{\pi}^2}\{
    F_{Th}^{\gamma\pi}+\frac{m_{\pi}\omega^2}{\alpha}\frac{\bar{\alpha_{\pi}}^2(1+\cos^2\theta)+\bar{\beta_{\pi}}^2 \cos\theta}
    {(1+\frac{\omega}{m_{\pi}} \cos\theta)^3}\}
\end{equation}

\noindent $\bar{\alpha_{\pi}}$ and $\bar{\beta_{\pi}}$ are the
electric and magnetic polarizabilities, respectively. Theoretically
polarizabilities can be described in the framework of chiral
perturbation theory, an effective field theory derived from QCD
where mesons are the relevant degrees of freedom. The low energy
expansion parameter is $\frac{m_{\pi}}{p}$ and higher order terms
$p^n$ contain low energy constants derived from experiment. By now
the calculations are of very high accuracy and constitute a
challenge to experiments. While the first experiments using the same
technique of Primakoff scattering obtained a value of
$\bar{\alpha_{\pi}}=(6.8\pm1.4_{stat.}\pm1.2_{sys})\cdot10^{-4}fm^3$
\cite{Serpukhov_pol} a more recent extraction using threshold pion
production in electron scattering off a proton obtained
$\bar{\alpha_{\pi}}-\bar{\beta_{\pi}}=(11.6\pm0.5_{stat.}\pm3.0_{sys}\pm0.5_{model})\cdot10^{-4}fm^3$
\cite{Mami_pol}. These numbers can be compared directly if one
assumes $\bar{\alpha_{\pi}}+\bar{\beta_{\pi}}=0$
\par
COMPASS has taken a first data set and within a few days of running
about 35000 Primakoff events have been recorded with a 4-momentum
transfer $t\leq 0.0015 (GeV/c)^2$, which separates the Primakoff
region from the diffractive background dominating a larger values of
t. We expect to improve in statistical accuracy by about a factor
4-5 over the previous experiment from Serpukhov.
\par
\begin{figure}
  \includegraphics[width=2.0in]{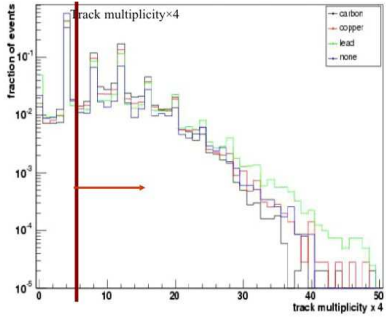}\\
  \caption{Track multiplicities (x4) as observed in the COMPASS silicon telescope downstream
  of the target using incoming pions. This info is used in a fast online data filter.}
  \label{multiplicities}
\end{figure}

In addition, we have recorded events with neutral pions in the final
state to obtain a new value for the chiral anomaly using the
Primakoff reaction $\pi^-\rightarrow\pi^-\pi^0$. In the same data
set we also have obtained events for diffractive production off a
nucleus. Using a CH-target and a multiplicity trigger (fig.
\ref{multiplicities}) we investigated reactions like
$\pi^-p\rightarrow\pi^-\pi^-\pi^+p$ as well as
$\pi^-p\rightarrow\gamma\gamma\pi^-p$. This programm is a
continuation of the VES \cite{VES} and BNL \cite{BNL} measurements
on diffractive production on exotic mesons, however, with much
higher beam energy.

\subsection{Future of meson spectroscopy using central production}
\begin{figure}
  \includegraphics[width=1.75in]{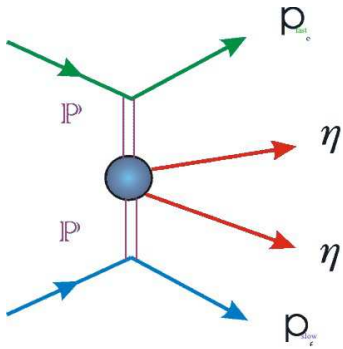}\\
  \caption{Sketch of central production of a state X decaying into $\eta\eta$.}\label{central_production}
\end{figure}

Central production ($pp\rightarrow p_{fast}p_{slow}~X$) of a state X
has proven to be a very useful tool in the search for exotic mesons
and gluon rich states \cite{WA102} (fig. \ref{central_production}) .
As the cross section drops very fast with the mass of X we need high
beam intensities and acceptance to extend the measurements into the
region above 2 $GeV/c^2$. Using a 270 GeV/c proton beam, a liquid
hydrogen target and full electromagnetic calorimetry we will
investigate this reaction in 2007. Beam intensities of a few
$10^7$/spill are required to obtain a major gain in statistical
significance as compared to previous experiments. This experimental
challenge asks for fast triggering ($p_T-balance$, ECAL) and high
resolution.

\section{Summary}
Using data taken in 2002-2003 COMPASS has made a major step forward
in determining the gluon polarization in a direct measurement on
polarized deuterons. The results obtained using high $p_T$ hadron
pairs points to a very small value for $\Delta G/G$ at $x_G\approx
0.1$ which does not favor expectations derived from QCD analysis of
polarized DIS on nucleons. No significant sign for transversal
asymmetries has been found so far, neither in transversity nor for
the Sivers functions.
\par
First data for the measurement of the pion polarizability has been
collected and are currently being analyzed. This marks the beginning
of the physics program with hadron beams within COMPASS, which is
planned to be continued in 2007 with measurements on central
production of hadrons.






\bibliographystyle{aipproc}
\bibliography{sample}

\begin{thebibliography}{99}
\bibitem{COMPASS_A1}E.S. Ageev et al., Phys.Lett.B612:154-164,2005
\bibitem{COMPASS_pt}E.S. Ageev et al., submitted to Phys.Lett.B
\bibitem{SMC_DeltaG}B. Adeva et al., Phys.Rev.D70:012002,2004
\bibitem{HERMES_pt}A. Airapetian et al., Phys.Rev.Lett.84:2584,2000
\bibitem{DeltaG_theory}M. Gluck, E. Reya, M. Stratmann, W. Vogelsang, Phys.Rev.D63:094005,2001
\bibitem{COMPASS_transversity}V.Yu.Alexakhin et al. Phys.Rev.Lett.94:202002,2005
\bibitem{HERMES_transversity}A. Airapetian et al.,
Phys.Rev.Lett.94:012002,2005
\bibitem{Collins}J. Collins, Nucl.Phys.396:161,1993
\bibitem{Sivers}D. Sivers, Phys.Rev.D.41:83,1990
\bibitem{Serpukhov_pol}Yu.M. Antipov et al., Phys.Lett.B121:445,1983
\bibitem{Mami_pol}J. Ahrens et al., Eur.Phys.J.A23:113-127,2005
\bibitem{VES} D.V. Amelin et al. Phys.Lett.B356:595-600,1995
\bibitem{BNL} E.I. Ivanov et al. Phys.Rev.Lett.86:3977-3980,2001
\bibitem{WA102}D. Alde et al., Phys.Lett.B205:397,1988  and D. Barberis et al.,
Phys.Lett.B397:339-344,1997
\end{thebibliography}


\IfFileExists{\jobname.bbl}{}  {\typeout{}
\typeout{******************************************}
\typeout{** Please run "bibtex \jobname" to optain}
\typeout{** the bibliography and then re-run LaTeX}
\typeout{** twice to fix the references!}
\typeout{******************************************}  \typeout{}  }

\end{document}